\documentclass[aps,pra,twocolumn,superscriptaddress]{revtex4}
\usepackage{amsfonts}
\usepackage{amssymb}
\usepackage{amsmath}
\usepackage{epsfig}
\usepackage{color}
\usepackage{graphics, graphicx}
\usepackage{bbold}
\usepackage{psfrag}
\usepackage{mathcomp}
\usepackage{subfigure}
\usepackage{verbatim}
\usepackage{CJK}
\usepackage[colorlinks,citecolor=blue]{hyperref}

\begin{document}


\title{Electron Teleportation via Multiple Majorana Bound States in a Superconductor Island}


\author{Zhen-Tao Zhang}
\email[Corresponding author:]{zhangzhentao@lcu.edu.cn}

\author{Bao-Long Liang}
\author{Zhen-Shan Yang}

\affiliation{School of Physics Science and Information Technology, Shandong Key Laboratory of Optical Communication Science and Technology, Liaocheng University, Liaocheng 252059, China}


\date{\today}

\begin{abstract}
Electron teleportation via two separate Majorana bound states(MBSs) is a manifestation of the non-locality of MBSs. A superconductor may host multiple separate or partial overlapping MBSs, and it is difficult to distinguish them. Here, we have studied the electron teleportation between two quantum dots via multiple MBSs in a superconductor island, two of which couple with the quantum dots. We find that in the absence of Majorana coupling, both elastic and inelastic electron transfers are allowed for specific system settings, and the extent to which the island state is changed after the teleportation relies on the initial state of the MBSs. In the presence of Majorana couplings, the elastic and inelastic teleportations are selective according to which pair of MBSs are coupled. Meanwhile, the cotuneling processes are distinct for different MBSs coupling types. In addition, we have investigated the effect of the asymmetry of the tunnelings to quantum dots on the transport. Our findings are meaningful for resolving transport signatures induced by topological MBSs and that stems from  nontopological quasiparticle.
\end{abstract}


\maketitle

\section{Introduction}
Majorana bound states in condensed matter system are exotic quasiparticles which exhibit non-Abelian statistic property. Due to their fundamental importance and potential applications in topological quantum computation \cite{Kitaev01,Ivanov01,Bonderson08,Nayak08,Flensberg11,Alicea11,Hassler11,Zhang13,Hyart13,Knapp16,Aasen16,Karzig19,Zhang19,Knapp20,Liu21}, a large number of systems are constructed to produce MBSs. These include topological insulator proximited with a superconductor \cite{Fu08,Xu14}, semiconductor nanowires-superconductor heterostructures \cite{Lutchyn10,Oreg10,Lutchyn18,Cao23}, magnetic atoms chains \cite{Nadj-Perge13} and Iron based superconductor \cite{Wang15,Zhang18,Liu23}. Although several phenomena in the candidate systems are consistent with the MBSs interpretation \cite{Deng12,Deng16,Das12,Mourik12,Nadj-Perge14,Nichele17,Rokhinson12,Wang22}, it is not completely clear whether these phenomena come from MBSs or nontopological subgap states \cite{Chen17,Suominen17,Liu17,Pan20,Yu21}, such as Andreev bound states. Therefore, more unique signatures of MBSs are desirable for confirming their existence. \\
\indent Nonlocality of MBSs can be used to distinguish them from local subgap states with topological trivial origins. Two MBSs form a normal Fermion, thus, their state could be distributed spatial nonlocally. Ref. \cite{Tewari08} predicted a nonlocal electron teleportation process between two dots mediated by a p-wave superconductor. However, the teleportation process is accompanied with crossed-Andreev-reflection-induced Cooper pair splittings, which blurs the transport signature. To address this problem, Ref. \cite{Fu10} suggested to use the charging energy of a mesoscopic superconductor to dismiss crossed Andreev reflections. \\
\begin{figure}
	\includegraphics[width=8.5cm]{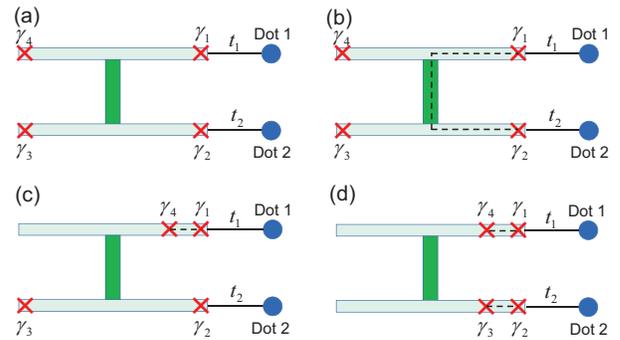}
	\caption{Sketches of Majorana island coupled with quantum dot 1 and 2. The island consists of two superconductor-proximited nanowires connected by a superconductor. Under certain conditions, the nanowires would host four separate or partial overlapping MBSs. The right two MBSs $\gamma_1, \gamma_2$ are tunneling coupled to Dot 1 and Dot 2, respectively. (a) the ideal case with four  separate MBSs. (b) the case with interwire Majorana coupling. (c), (d) the cases with innerwire Majorana couplings.  \label{fig1}}
\end{figure}
\indent Recently, semiconductor nanowires proximited with a floating topological superconductor island are frequently concerned, with the hope of characterizing transport properties of MBSs \cite{Albrecht16,Heck16,Chiu17,Michaeli17,Lutchyn17,Li20,Liu20,Whiticar20,Thamm21,Shen21,Lai21,Vayrynen21,Abboud22} or building topological qubits immune from quasiparticle poisonings \cite{Plugge17,Karzig17,Steiner20,Munk20,Gau20,Karzig21,Zhang21}. Generally, in order to detect the nonlocality of MBSs and encode topological qubit, multiple MBSs are involved in the system. Lately, Ref. \cite{Ekstrom20,Nitsch22,Hao22} studied transport properties between terminals through two pairs of MBSs in the Coulomb blockade regime. Most recently, Ref. \cite{Souto22} discussed the approach of multiterminal transport spectroscopy to distinguish local and nonlocal subgap states in a superconducting island. It is natural to ask whether the electron teleportation mediated by multiple MBSs is spatially local or nonlocal, and how the initial state of MBSs and couplings between MBSs affect the electron transport.\\
\indent To end this, we investigate the electron transfer between two quantum dots via semiconductor nanowires on a superconductor island, which could host four MBSs. Each dot is tunnel coupled with a MBS. We find that in the absence of Majorana couplings the transport process is dependent on the initial state of the MBSs. When the parity of the dot-coupled MBSs is disentangeled with that of the other MBSs, an electron could be teleported between the dots with the state of the island unchanged. On the other side, if these two pairs of MBSs are entangled, the state of the island would alter after the transportation. In the presence of Majorana couplings, the transport behavior would be modified according to the coupling type. The interwire Majorana couplings enable elastic electron transfers and cotunneling processes. Contrarily, the innerwire couplings could support inelastic tunneling, and suppress the electron cotunneling as the island is in even subspace. In addition, we have studied the effect of asymmetric couplings between MBSs and the dots on the transport.\\
\indent The paper is organized as follows. In Sec. 2, we introduce the model and the Hamiltonian of the quantum dots-Majorana island system. In Sec. 3, we calculate and analyze the electron transport and the state evolution of the system without Majorana couplings. Sec. 4 is devoted to investigate the impact of Majorana couplings on the electron sequential tunnelings and cotunnlings. In Sec. 5, we revisit the electron transfer process with asymmetric tunneling couplings. A conclusion is given in the last section.

\section{DQD-Majorana island system}\label{sec2}
We consider a floating superconductor island proximited with two parallel nanowires. Under certain conditions, the nanowires would host four MBSs at their ends, marked as $\gamma_i (i=1,2,3,4)$, as shown in Fig. 1. $\gamma_1$ and $\gamma_2$ are tunnel coupled to quantum dot 1 and dot 2, respectively. The Hamiltonian of the quantum dots-island system reads $H=H_d+H_c+H_T+H_{int}+H_{inn}$, where $H_d=\sum_{i=1,2}\epsilon_i\hat{d}_i^\dagger\hat{d}_i$ is Hamiltonian of the two quantum dots, and $H_c=E_C(\hat{N}-N_g)^2$ represents the charging energy of the island with $E_C=e^2/2C$. $\hat{N}$ and $N_g$ denote the total charge number and charge offset in the island, respectively. Here, $N_g$ is set at 1/2mod2, where the charging energy of the state $N=0(mod2)$ and that of $N=1(mod2)$ are equal.  $H_{int}=iJ(\gamma_1\gamma_2+\gamma_3\gamma_4)/2$ describe couplings between MBSs locating at different wires, and $H_{inn}=iJ'(\gamma_1\gamma_4+\gamma_2\gamma_3)/2$ denotes innerwire MBSs couplings. Finally, the tunneling between the quantum dots and the superconducting island is described by the Hamiltonian 
\begin{equation}
	H_T=(t_1\hat{\gamma}_1\hat{d}_1+t_2\hat{\gamma}_2\hat{d}_2)e^{i\hat{\phi}/2}+\text{h.c.},
\end{equation}
where $t_i$ is the tunneling amplitude between mode $\hat{\gamma}_i$ and dot $i$, and $\hat{\phi}$ is the phase of the island with $[\hat{\phi}, \hat{N}]=2$. $\hat{d_1},\hat{d_2}$ are annihilation operators of dot 1 and dot 2. Here, we have assumed that each dot only couples to its nearest MBS. In the regime $E_C\gg J,J'$, the states with $N=N_g\pm\frac{1}{2}$ are preferable in energy, thereby, the island would be constrained in this subspace. As a consequence, Cooper-pair tunnelings between the island and the dots are inhibited, and only single electron tunnelings could occur.\\
\indent Because two MBSs form a Dirac Femion, we can make a transformation from Majorana basis to Dirac basis. The combination of four MBSs has a selection freedom. According to the couplings between MBSs, we introduce two Dirac bases for further calculation. The first transformation is 
\begin{eqnarray}
	\hat{f_1}=\frac{1}{2}(\hat{\gamma}_1+i\hat{\gamma}_2),\qquad \hat{f_1}^\dagger=\frac{1}{2}(\hat{\gamma}_1-i\hat{\gamma}_2),\\ 	
	\hat{f_2}=\frac{1}{2}(\hat{\gamma}_3+i\hat{\gamma}_4),\qquad \hat{f_2}^\dagger=\frac{1}{2}(\hat{\gamma}_3-i\hat{\gamma}_4).
\end{eqnarray}

Therefore, the charge operator can be written as $\hat{N}=\hat{n}_1+\hat{n}_2+2N_C$, where $\hat{n}_i=\hat{f}_i^\dagger \hat{f}_i\,(i=1,2)$ denote the occupation operator of the fermion mode $\hat{f}_i$, and $N_C$ is the number of Cooper pairs in the island. The state of the Majorana island can be described in the basis of $|n_1,n_2, N_c\rangle$. It is worth to note that while mode $f_1$ is coupled both to dot 1 and dot 2, the mode $f_2$ has no correlation with the two dots.\\
\indent The other basis transformation is
\begin{eqnarray}
	\hat{f_u}=\frac{1}{2}(\hat{\gamma}_1+i\hat{\gamma}_4),\\ 	
	\hat{f_d}=\frac{1}{2}(\hat{\gamma}_2+i\hat{\gamma}_3).
\end{eqnarray}
In this new basis, the mode $\hat{f_u}$ and $\hat{f_d}$ are coupled to dot 1 and dot 2, respectively. The state of the island can be represented using the basis of $|n_u,n_d, N_c\rangle_N$. The subscript $N$ is added to distinguish it from the first basis. \\

\section{electron transport: the case without Majorana couplings}
 In this section, we investigate the electron transport in the situation $J=J'=0$. Firstly, we assume the energy levels in the two quantum dots are both aligned with the chemical potential of the superconductor, i.e., $\epsilon_1=\epsilon_2=0$. Without losing generality, we suppose that the island stays within the even parity subspace before coupling to the quantum dots, i.e., $\{|00,N_C\rangle,|11,N_C-1\rangle\}$, and the quantum dots are initially prepared at the state $|n_{d1}=d^\dagger_1d_1=1, n_{d2}=d^\dagger_2d_2=0\rangle$. The state of the whole system can be expressed on the combined basis $|n_{d1}n_1n_2n_{d2},N_C\rangle$. Due to charge conservation of the whole system and the charge energy, the state is restricted in the subspace $\{|1000,N_C\rangle,\, |0100,N_C\rangle,\,|0001,N_C\rangle,\, |1110,N_C-1\rangle,\,|0010,N_C\rangle,\,|0111,N_C-1\rangle\}$.
 Under this basis, the Hamiltonian can be written as
 \begin{equation}\label{eq6}
 	H=
 	\begin{pmatrix}{}
 	0   & t_1^*   & 0     & 0   & 0       & 0\\
 	t_1 & 0       & it_2  & 0   & 0       & 0\\
    0   & -it_2^* & 0     &	0   & 0       & 0\\      
 	0   & 0       & 0     & 0   & t_1^*   & 0 \\    
 	0   & 0       & 0     & t_1 & 0       & -it_2 \\  
 	0   & 0       & 0     & 0   & it_2^*  & 0    
 	\end{pmatrix}.
 \end{equation}
 We can see that the Hamiltonian is block-diagonalized with respect to the subspaces $\{|1000,N_C\rangle, \,|0100,N_C\rangle,\,|0001,N_C\rangle\}$ and $\{|1110,N_C-1\rangle,\,|0010,N_C\rangle,\,|0111,N_C-1\rangle\}$. For simplicity, we consider the situation $t_1=it_2=t$ with $t$ be real. 
 If the initial state is $|1000,N_C\rangle$, the probability amplitude of observing an electron at the dot 2 at a later time T is given by
 \begin{equation}
 	\langle0001,N_C|e^{-iHT}|1000,N_C\rangle=\frac{1}{2}(\cos(\sqrt{2}tT)-1).
 \end{equation}
We have calculated the state evolution of the system according to the Hamiltonian \ref{eq6}, as shown in Fig. 2. We can see that the coherent couplings between the island and the dots can transfer an electron back and forth between the two dots periodically via the intermediate state $|0100,N_C\rangle$. At the time $T=\sqrt{2}\pi/{2t}$ an electron is transported to dot 2. The coherent transport processes can be illustrated as following:
\begin{eqnarray}
	|1000,N_C\rangle\rightleftarrows	|0100,N_C\rangle\rightleftarrows	|0001,N_C\rangle.\label{tp1}
\end{eqnarray} 
Similarly, if the initial state is $|1110,N_C\rangle$, the probability amplitude of observing an electron at the dot 2 at a later time T is given by
\begin{equation}
	\langle0111,N_C-1|e^{-iHT}|1110,N_C-1\rangle=\frac{1}{2}(1-\cos(\sqrt{2}tT)).
\end{equation}
\begin{figure}
\includegraphics[width=8.5cm]{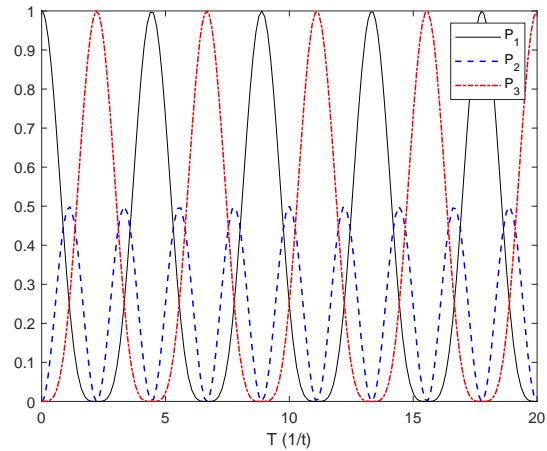}
	\caption{State evolution without Majorana couplings. The initial state of the island-quantum dots system is $|1000,N_C\rangle$. The populations of $|1000,N_C\rangle$ ($P_1$, black solid line), $|0100,N_C\rangle$ ($P_2$, blue dashed line), $|0001,N_C\rangle$ ($P_3$, red dash-dotted line) are calculated numerically according to the Hamiltonian \ref{eq6}. The parameters are set as: $t_1=it_2=t, \,\epsilon_1=\epsilon_2=0$.\label{fig2} }
\end{figure}
The related transport process can be illustrated by:
\begin{eqnarray}
	|1110,N_C-1\rangle\rightleftarrows	|0010,N_C\rangle\rightleftarrows	|0111,N_C-1\rangle.\label{tp2}
\end{eqnarray}
  For the above two cases, when an electron is transferred to dot 2, the island is recovered to its initial state. Therefore, both the process \ref{tp1} and the process \ref{tp2} are elastic sequential electron tunnelings. On the other hand, the main difference between them lies at that the latter takes place via cross Andreev reflections, involving Cooper-pair combination and breaking in the island, while the former is realized via normal resonant tunnelings. \\
 \indent Now we consider a generic initial state. The island is at a superposition state $a|00,N_C\rangle+b|11,N_C-1\rangle$, and the dots are prepared at $|n_{d1}=1,n_{d2}=0\rangle$. In this case, the initial state of the whole system is  $|\psi(0)\rangle=a|1000,N_C\rangle+b|1110,N_C-1\rangle$. We can obtain the possibility of an electron occupying dot 2:
 \begin{eqnarray}
 	&|\langle0001,N_C|e^{-iHT}|\psi(0)\rangle|^2+|\langle0111,N_C-1|e^{-iHT}|\psi(0)\rangle|^2\nonumber\\
 	&=\sin^4(\frac{\sqrt{2}tT}{2}),
 \end{eqnarray} 
which is independent on the initial state. At $T=\sqrt{2}\pi/{2t}$ the electron would be transported to dot 2 as same as the case with initial state $|1000,N_C\rangle$ or $|1110,N_C-1\rangle$. However, the state of the island is changed to $a|00,N_C\rangle-b|11,N_C-1\rangle$ up to a whole phase factor. Specially, if $|a|^2=|b|^2=1/2$, the state of the island at $T=\sqrt{2}\pi/{2t}$ is orthogonal to its initial state. Therefore, these transport events are inelastic sequential tunnelings. To interpret this transition, we make use of the alternative basis $|n_u,n_d,N_C\rangle_N$.
 Suppose the island initially prepared at 
the state $(|00,N_C\rangle+|11,N_C-1\rangle)/\sqrt{2}$ in the basis $|n_1n_2,N_C\rangle$. Transforming to the basis $|n_un_d,N_C\rangle_N$, the state is of the form $|00,N_C\rangle_N$. Using the new basis, the electron transport process is taken as following:
\begin{equation}
	|1000,N_C\rangle_N\rightleftarrows	|0100,N_C\rangle_N\rightleftarrows	|0111,N_C-1\rangle_N,\label{tp3}
\end{equation}
It can be seen that when an electron populates dot 2, the state of the island becomes $|11, N_C-1\rangle_N$, which is orthogonal to its initial state $|00,N_C\rangle_N$. From the above results, we can conclude that the efficiency of the electron transport is independent on the initial state of the island, however, the state evolution of the island during a transport period relies on its initial state. \\

\section{electron transport: the case with Majorana couplings}
Now we investigate the impacts of Majorana couplings on the electron transport between the two dots. Both the interwire and innerwire Majorana couplings could lift the degeneracies of the island and thus modify the tunneling process. For clarity, we would study their effects exclusively in the following.
\begin{figure}
	\includegraphics[width=8.5cm]{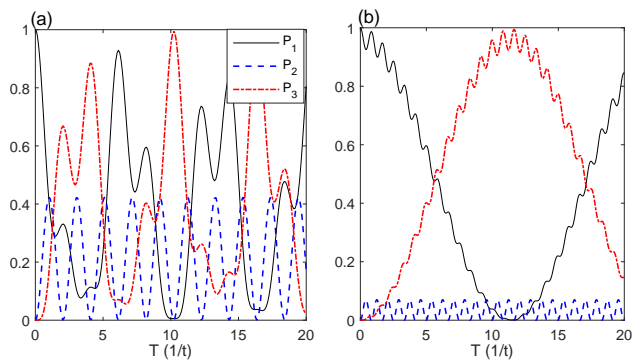}
	\caption{ State evolution in presence of interwire Majorana couplings. The initial state is $|1000,N_C\rangle$, and the populations of $|1000,N_C\rangle$ ($P_1$, black solid line), $|0100,N_C\rangle$ ($P_2$, blue dashed line), $|0001,N_C\rangle$ ($P_3$, red dash-dotted line) are calculated according to the Hamiltonian \ref{hs}. (a) $J=1.2t_1$; (b) $J=7t_1$. The other parameters are: $t_1=it_2=t, \,\epsilon_1=\epsilon_2=0$.\label{fig3}}
\end{figure}
\subsection{interwire Majorana couplings}
The interwire Majorana coupling may be happened between $\gamma_1$ ($\gamma_3$) and $\gamma_2$ ($\gamma_4$). The interaction Hamiltonian reads
\begin{equation}
	H_{int}=i\frac{J}{2}(\gamma_1\gamma_2+\gamma_3\gamma_4)=J(\hat{f_1}^\dagger\hat{f_1}+\hat{f_2}^\dagger\hat{f_2}-\frac{1}{2}),
\end{equation}
with $J$ be the coupling strength. The occupation operators $\hat{n}_1$, $\hat{n}_2$ are conserved quantities respected by $H_{int}$, therefore we choose the basis $|n_1,n_2,N_C\rangle$ for the island. Consequently, the subspace $\{|1000,N_C\rangle, \,|0100,N_C\rangle,\,|0001,N_C\rangle\}$ and $\{|1110,N_C-1\rangle,\,|0010,N_C\rangle,\,|0111,N_C-1\rangle\}$ are decoupled and permit similar tunneling processes. Thus, we focus on the dynamical behavior in the former.\\
\indent We assume the initial state of the whole system is $|1000,N_C\rangle$. The Hamiltonian in the related subspace can be written as
 \begin{equation}
 	H_s=
 	\begin{pmatrix}{}
 		\epsilon_1       & t_1^*          & 0          \\
 		t_1              & J              & it_2       \\
 		0                & -it_2^*        & \epsilon_2     
 		\end{pmatrix}.\label{hs}
 \end{equation}
When the resonance condition $\epsilon_1=\epsilon_2=J$ is met, the electron tunneling is the same as the case with $\epsilon_1=\epsilon_2=J=0$. In this case, the elastic sequential tunnelings occur: when the electron is transferred to dot 2, the MBSs recover to their initial state. It is worth to note that the resonance condition in the other subspace is different, therefore sequential tunnelings would not coexist in the two subspace if $J$ is much larger than the tunneling rates $t_1,t_2$.\\
\indent We investigate the transition from sequential tunneling to elastic cotunneling of the system by setting $\epsilon_1=\epsilon_2=0$. The states of the system as a function of time are calculated numerically for two different coupling strengths, as shown in Fig. 3. When $J=1.2t_1$ (shown in Fig. 3a), the population of $|0100,N_C\rangle$ oscillates periodically with an amplitude smaller than that without Majorana coupling. At the same time, the possibilities of the states $|1000,N_C\rangle$ and $|0001,N_C\rangle$ are both varying in a beating-like manner. As $J=7t_1$ (shown in Fig. 3b), the state $|0100,N_C\rangle$ is barely populated due to the large detuning between $|1000,N_C\rangle$ and $|0100,N_C\rangle$. Meanwhile, the other two states are oscillating out of phase with maximum amplitude, both in a beating manner. The results tell us that when the interwire Majorana coupling is much larger that $t_1,t_2$, the electron in dot 1 could elastically cotunnel to dot 2, leaving the state of the MBSs unchanged. Actually, the electron cotunneling also exist in odd subspace.\\
\indent  It is worth to note that the coupling $\gamma_1$ and $\gamma_2$ behave as a delocalized bound state coupled both to dot 1 and dot 2. Therefore, in the low energy regime ($\epsilon_1,\epsilon_2$ is smaller than superconductor gap), the interwire Majorana coupling would suppress inelastic tunnelings, and only allows elastic tunnelings. 
\subsection{innerwire Majorana couplings}
Now we turn to the effect of innerwire Majorana couplings on the transport process. For simplicity, we assume that $\gamma_1$($\gamma_2$) couples to $\gamma_4$($\gamma_3$) with the same strength $J'$. The interaction Hamiltonian can be written as
\begin{equation}
   	H_{inn}=i\frac{J'}{2}(\gamma_1\gamma_4+\gamma_2\gamma_3)=J'(\hat{f_u}^\dagger\hat{f_u}+\hat{f_d}^\dagger\hat{f_d}-1).
\end{equation} 
Obviously, the occupation operators $\hat{n}_u$ and $\hat{n}_d$ commute with the above Hamiltonian, and the states $|n_un_d,N_C\rangle_N$ are eigenstate of the island if $t_1=t_2=0$. Therefore, we should use the basis $|n_{d1}n_un_dn_{d2},N_C\rangle_N$ for describing the state of the system. In the following, we would discuss the transport process with an initial state in even or odd subspace of the island.\\
\indent In the even subspace of the island, the degeneracy of $|00,N_C\rangle_N$ and $|11,N_C-1\rangle_N$ is lifted with energy splitting $2J'$. Started with the state $|1000,N_C\rangle_N$, the sequential tunneling process $	|1000,N_C\rangle_N\longrightarrow	|0100,N_C\rangle_N\longrightarrow|0111,N_C-1\rangle_N$ would take place when the resonant condition $\epsilon_1=-\epsilon_2=J'$ is met. Similarly, started with the state $|1110,N_C-1\rangle_N$, the sequential tunneling process $|1110,N_C\rangle_N\longrightarrow	|0010,N_C\rangle_N\longrightarrow|0001,N_C-1\rangle_N$ would occur as $\epsilon_1=-\epsilon_2=-J'$. It can be seen that these two processes would not happen with the same parameter configuration since the resonant conditions are distinct. Thus, the sequential tunnelings are inelastic. Moreover, if $J'\gg t_1,t_2$, the cotunneling is suppressed due to the unequal energies between $|00,N_C\rangle_N$ and $|11,N_C-1\rangle_N$. \\
\indent On the other hand, the basis states in the odd subspace $|01,N_C\rangle_N$ and $|10,N_C\rangle_N$ are degenerate due to the symmetry of the Majorana couplings. However, these two states lead to different sequential tunneling processes:
\begin{eqnarray}
	|1010,N_C\rangle_N\longrightarrow	|0110,N_C\rangle_N\longrightarrow|0101,N_C\rangle_N,\\
	|1100,N_C\rangle_N\longrightarrow	|0000,N_C+1\rangle_N\longrightarrow|0011,N_C\rangle_N.
\end{eqnarray}   
The corresponding resonant conditions are $\epsilon_1=\epsilon_2=J'$ and $\epsilon_1=\epsilon_2=-J'$, respectively. Since then, the above two tunneling processes take place exclusively. Furthermore, in the regime $J'\gg t_1,t_2$, an electron could cotunnel from dot 1 to dot 2 if the island is in odd subspace. Explicitly, the transition $	|1010,N_C\rangle_N\leftrightarrow|0101,N_C\rangle_N$ could happen through a  virtual excitation to the state $|0110,N_C\rangle_N$, as well as the cotunneling $|1100,N_C\rangle_N\leftrightarrow|0011,N_C\rangle_N$ via the state $|0000,N_C+1\rangle_N$. \\
\indent  From the above analysis, we can see that the innerwire Majorana couplings lift the sequential tunneling processes in each subspace, and only allow for inelastic tunneling. Moreover, the electron cotunneling could occur merely in the odd subspace. Thus, compared with the case of interwire Majorana coupling, the cotunneling is partially suppressed, which is consistent with the result of two local subgap states \cite{Souto22}. Therefore, the innerwire Majorana couplings make the MBSs act like two local subgap states locating in each nanowire.
   \begin{figure}
   	\includegraphics[width=9.5cm]{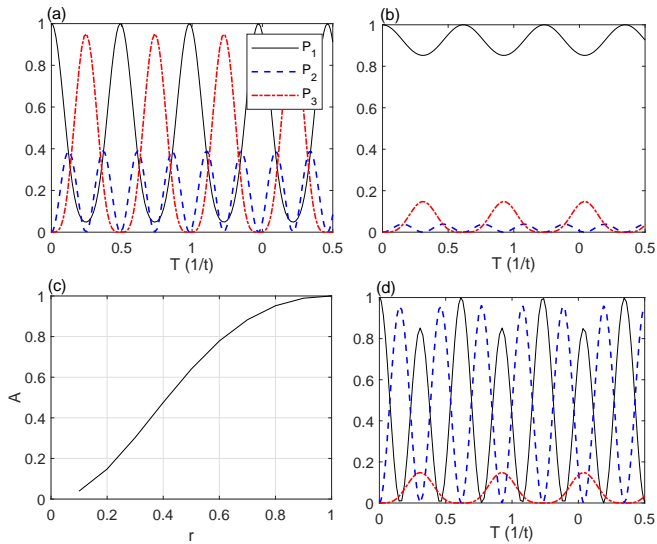}
   	\caption{Effect of asymmetric tunneling couplings. The state evolution starting with $|1000,N_C\rangle$ are plotted with the ratio $r=$0.8 (a), 0.2 (b), 5 (d). the populations of $|1000,N_C\rangle$ ($P_1$, black solid line), $|0100,N_C\rangle$ ($P_2$, blue dashed line), $|0001,N_C\rangle$ ($P_3$, red dash-dotted line) are calculated from Eq. \ref{p1}-\ref{p3}. (c) the amplitude $A$ of $P_3$ is plotted as a function of $r$. In (a)-(b), $t_1=rt,\,it_2=t$; in (d), $t_1=t,\,it_2=t/r$. The other parameters are set as: $J=J'=0$, $\epsilon_1=\epsilon_2=0$.\label{fig4}}
   \end{figure}
\section{effect of asymmetric tunnel couplings }
Before conclusion, we would concern how the transport be affected by asymmetric tunnelings from the island to dot 1 and dot 2, ie., $|t_1|\neq |t_2|$. For clarity, we assume the couplings between MBSs are vanishing in this section. Under this condition, the initial state has no effect on the efficiency of the electron transport. We choose the initial state be 
$|1000\rangle$ without losing generality. The state of the system would localized in the subspace $\{|1000,N_C\rangle, \,|0100,N_C\rangle,\,|0001,N_C\rangle\}$. The Hamiltonian is with the form of 
 \begin{equation}
	H_s=
	\begin{pmatrix}{}
		0             & t_1^*          & 0          \\
		t_1           & 0              & it_2       \\
		0             & -it_2^*        & 0     
	\end{pmatrix}.\label{hs1}
\end{equation}
Through some algebraic calculation, we can get the populations of the states $|1000,N_C\rangle, \,|0100,N_C\rangle,\,|0001,N_C\rangle$ at $T$ are 
\begin{eqnarray}
&&P_1=\frac{[|t_2|^2+|t_1|^2\cos(DT)]^2}{D^4},\label{p1}\\
&&P_2=\frac{|t_1|^2\sin^2(DT)}{D^2},\label{p2}\\
&&P_3=\frac{4|t_1t_2|^2\sin^4\frac{DT}{2}}{D^4}\label{p3}
\end{eqnarray} 
respectively, where $D=\sqrt{|t_1|^2+|t_2|^2}$. We can see that $P_3$ oscillates with amplitude 
\begin{equation}
	A=\frac{4|t_1t_2|^2}{D^4}=\left(\frac{2}{\frac{|t_1|}{|t_2|}+\frac{|t_2|}{|t_1|}}\right)^2,
\end{equation}
which is determined by the ratio $r=|t_1|/|t_2|$. When $|t_1|=|t_2|$, the oscillating amplitude is equal to 1, which has been shown in Fig. 2. \\
\indent To explain the effect of asymmetric tunnelings, we have calculated numerically the time evolution of $P_1,P_2,P_3$ for three combinations of $t_1$ and $t_2$ (shown in Fig. 4). In Fig. 4(a) and (b), the ratio $r$ is set to be 0.8 and 0.2 respectively. We can see that when $r=0.8$, $P_3$ oscillates with an amplitude 0.95. As $r=0.2$, the oscillation of $P_3$ is largely suppressed with $A=0.15$. We have also plotted the amplitude $A$ as a function of $r$ in Fig.4(c) on the range $0.1\leq r\leq1$, according to Eq. 22. It can be seen that the amplitude decreases slowly with the declining of $r$ near the symmetry point $r=1$. In Fig. 4(d), we illustrate the populations with the values of $t_1$ and $t_2$ reversed relative to Fig. 4(b), ie., $r=5$ and the value of $D$ is unchanged. We can see $P_3$ has no difference with the case $r=0.2$, as expected from Eq. 22. However, the other populations behaves distinctively compared to Fig. 4(b), which can be interpreted by Eq. 19 and 20. \\
\indent Based on the above results, we can remark that a weak asymmetry of $t_1$ and $t_2$ has little influence on the coherent electron transport. However, if the ratio $r$ deviates far from 1, the transport process between the two dots would be significantly blocked.
\section{conclusion}
 We have studied the electron teleportation via a  superconductor island with four MBSs. In the absence of Majorana coupling, both elastic and inelastic electron transfers are allowed for same system parameters, and the change of the island state in the teleport process only depends on the initial state of the MBSs. In the presence of Majorana couplings, the elastic and inelastic teleportations are selective according to the coupling type. Interwire Majorana couplings favor the elastic tunneling and enable the electron cotunneling. Innerwire Majorana couplings only allow the inelastic tunneling and partially suppress the cotunneling. In addition, the asymmetry of the tunnelings to quantum dots could lower the transfer efficiency.\\
 \indent Our results uncover that the electron teleportation via uncoupled multiple MBSs relies on the initial state of the MBSs. Furthermore, the transport property is sensitive to the MBSs couplings, which could make the system behave as local or nonlocal subgap states. This kind of sensitivity could be utilized to distinguish separate topological MBSs with overlaping MBSs or Andreev bound states.
\section{Acknowledgments}
 ZTZ is funded by Natural Science Foundations of Shandong Province of China (Grant No. ZR2021MA091), Introduction and Cultivation Plan of Youth Innovation Talents for Universities of Shandong Province (Research and Innovation Team on Materials Modification and Optoelectronic Devices at extreme conditions).






\end{document}